\documentclass[mypaper,8pt,twoside]{CoAst}
\usepackage{epsf,graphicx,fancyhdr,sfmath}
\renewcommand{\chaptermark}[1]{\markboth{#1}{}}

\newcommand{\kopf}{\small\itshape Comm.\ in Asteroseismology, N$^{\textsf{\underline{o}}}$ 159, 2009\\
Proceedings of the JENAM 2008 Symposium N$^{\textsf{\underline{o}}}$~4:
Asteroseismology and Stellar Evolution}

\newcommand{\Authors}[1]{\begin{center}\normalsize\bf\sf #1 \end{center}}

\renewcommand{\author}[1]{\begin{center}\normalsize\bf\sf #1 \end{center}}
\newcommand{\Address}[1]{\begin{center}\small\sf #1 \end{center}}

\newcommand{\Objects}[1]{{\vspace{3mm}\small \noindent Individual Objects: }\small\sf \hangindent=27truemm \hangafter=1 #1 \normalsize}

\renewenvironment{abstract}{\section*{Abstract}\normalsize\sf}{}
\newcommand{\References}[1]{\begin{flushleft}{\large References\\}\vspace*{2mm}\small #1 \end{flushleft}}

\newcommand{\chapterCoAst}[2]{\chapter[\sf\normalsize #1\\ \footnotesize \hspace*{5mm}by #2 \sf\normalsize][]{#1\\}\rhead[\fancyplain{}{\sf\footnotesize \center{#1}}]{\fancyplain{}{\sffamily\thepage}}\lhead[\fancyplain{\kopf}{\sffamily\thepage}]{\fancyplain{\kopf}{\sf\footnotesize \center{#2}}}}

\newcommand{\figureCoAst}[5]{\begin{figure}[#4]
\centering
\includegraphics*[#5]{#1}
\caption{#2}
\label{#3}
\end{figure}}

\renewcommand{\chaptername}{}

\newcommand{\Acknowledgments}[1]{\noindent\textbf{Acknowledgments.} #1}  

\def\rfr{\smallskip\par\noindent
        \hangindent=7truemm
        \hangafter=1}

\begin{document}
\sf
\chapterCoAst{Time-resolved spectroscopy of the planet-hosting sdB pulsator
              {V391\,Pegasi}}
{S.\,Schuh, R.\,Kruspe, R.\,Lutz, and R.\,Silvotti} 
\Authors{S.\,Schuh,$^{1}$ R.\,Kruspe,$^{1}$ R.\,Lutz,$^{1}$ and 
R.\,Silvotti\,$^{2}$} 
\Address{
$^1$ Institut f\"ur Astrophysik, Universit\"at G\"ottingen,
  Friedrich-Hund-Platz~1, 37077~G\"ottingen, Germany\\
$^2$ INAF - Osservatorio Astronomico di Capodimonte, via Moiariello
  16, 80131 Napoli, Italy
}
\noindent
\begin{abstract}
\label{ref:schuh}
The subdwarf~B (sdB) star V391~Peg oscillates in
short-period p modes and long-period g modes, making it one
of the three known hybrids among sdBs. As a by-product of the effort
to measure secular period changes in the p modes due to evolutionary
effects on a time scale of almost a decade, the O--C diagram has revealed
an additional sinusoidal component attributed to a periodic shift in
the light travel time caused by a planetary-mass companion around
the sdB star in a 3.2\,yr orbit. In order to derive the mass of the
companion object, it is necessary to determine the orbital
inclination. One promising possibility to do this is to use the
stellar inclination as a primer for the orbital orientation.
The stellar inclination can refer to the rotational or the pulsational
axis, which are assumed to be aligned, and can in turn then be derived
by combining measurements of $v_{rot}$ and $v_{rot}\sin{i}$. 

The former is in principle accessible through rotational
splitting in the photometric frequency spectrum (which has however not
been found for V391~Peg yet), while the projected rotational
velocity can be measured from the rotational broadening of spectral
lines. The latter must be deconvolved from the additional pulsational
broadening caused by the surface radial velocity variation in high
S/N phase averaged spectra. This work gives limits on pulsational radial
velocities from a series of phase resolved spectra.

Phase averaged and phase resolved high resolution echelle spectra were
obtained in May and September 2007 with the
9m-class Hobby-Eberly Telescope (HET), and one phase averaged spectrum in
May 2008 with the 10m-Keck~1 telescope\footnote{\sf %
  Data obtained with the Hobby-Eberly Telescope (joint project of U of Texas, Pennsylvania State
  U, Stanford U, U M\"unchen, U G\"ottingen) and the W.M.\ Keck
  Observatory (operated by CalTech, U of California,
  NASA), made possible by the generous financial support of the
  W.M.\ Keck 
  Foundation.%
}.
\end{abstract}

\Objects{V391\,Pegasi}
\section*{The hybrid pulsating sdB star V391~Pegasi and its
  planetary companion}
\figureCoAst{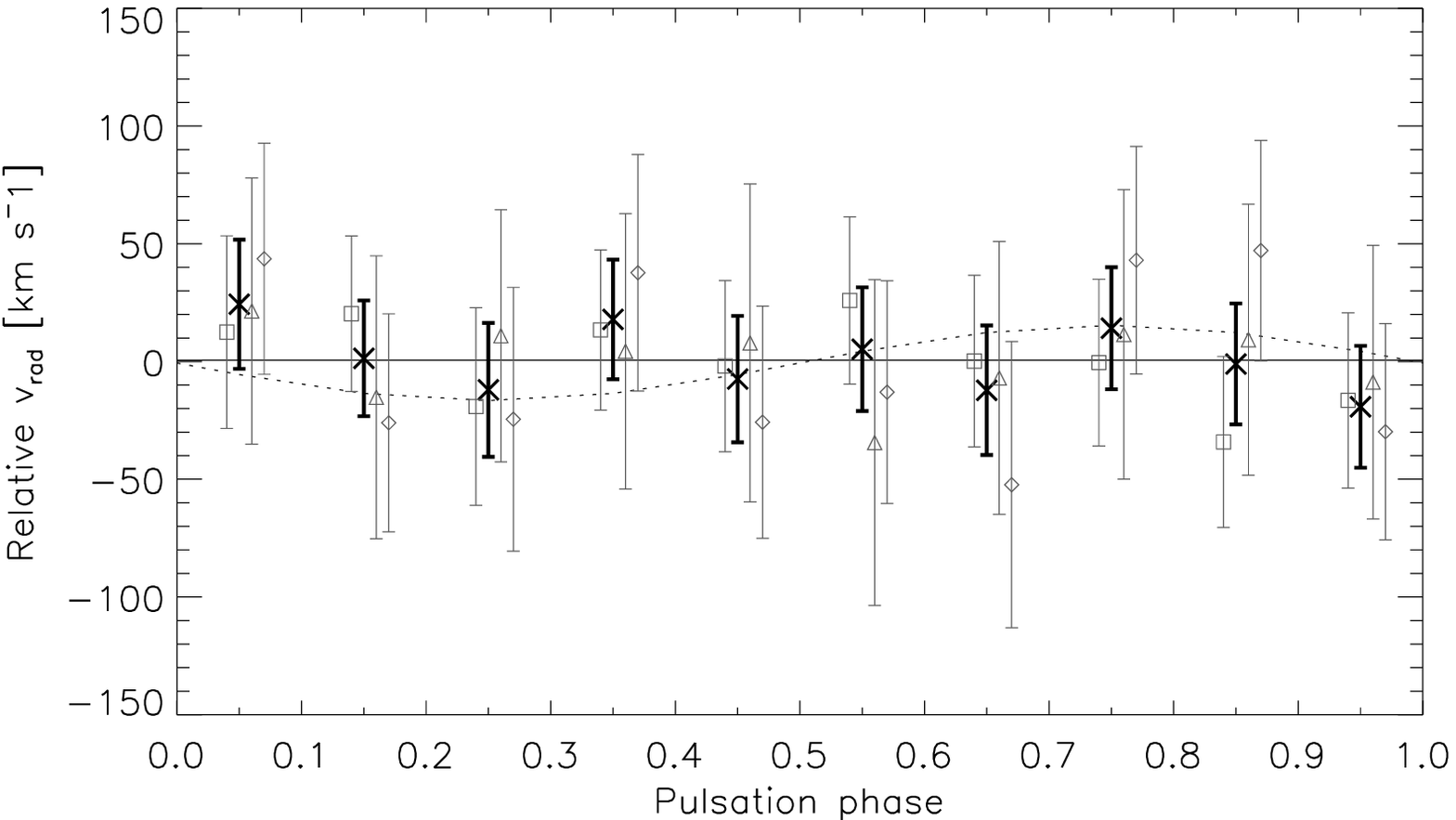}{Radial velocities derived for the H$\beta$
  (squares), H$\gamma$ and H$\delta$ (triangles and diamonds) lines
  from cross correlation with the template (with error bars, and
  plotted at a small phase offset for clarity); and weighted mean of
  the results (crosses and thick error bars) together with a constant
  and a sinusoidal 
curve.}{fig:schuh_fig1}{htb!}{clip,angle=0,width=0.76\textwidth} 
Subdwarf B stars (sdBs) are subluminous, evolved stars on the extreme
horizontal branch (EHB). They have a He burning core but, due to
previous significant mass loss, no H-shell burning in their thin
hydrogen shells. Their masses cluster around 0.5\,M$_{\odot}$.
Only a small fraction of the sdBs show pulsational variations, with
non-pulsators also populating the region in the HRD where the
pulsators are found. There are p (pressure) mode and g (gravity) mode 
types of pulsation. Three objects that show both 
mode types are referred to as hybrid pulsators and
among them is \mbox{V391 Peg} (HS\,2201+2610) which has five p modes 
({\O}stensen et al.\ 2001; Silvotti et al.\ 2002) and one g mode 
(Lutz et al.\ 2008, 2009). Silvotti et al.\ (2007)
detected parabolic and sinusoidal variations in the
observed--calculated (O--C) diagram constructed for the two main
pulsation frequencies at 349.5\,s and  354.2\,s over the observing
period of seven years. The sinusoidal component is attributed to the
presence of a very low-mass companion (\mbox{V391 Peg\,b},
$m\sin{i}=3.2 \pm 0.7\,$M$_{\sf{Jup}}$).  
The determination of the true mass of this 'asteroseismic planet' 
requires a constraint on the orbital inclination which
can presumably be determined via the stellar rotational inclination.
\section*{Limits on the pulsational radial velocities from phase
  resolved echelle spectra}
Echelle spectra of \mbox{V391 Peg} were taken during May and September
2007 with the HRS ($R=15\,000$) of the HET at the McDonald Observatory, 
and with HIRES ($R=31\,000$) at the Keck~1 telescope atop
Mauna Kea in May 2008. Data reduction was done with ESO-Midas using
standard procedures. Individual echelle orders were merged and the
final spectra were carefully normalized and finally summed (Kruspe
2009, diploma thesis, in prep.). This results in a set of individual spectra
($S/N \approx 3$), in particular two September 2007 high time
resolution series, and summed spectra for May and September 2007.
\par 
In our attempt to ``clean'' the relevant rotational broadening from
pulsational effects, the spectra in September obtained in time
resolved mode were combined to a series of ten phase resolved
averaged spectra ($S/N \approx 9$) for the main pulsation period of
349.5\,s (similar to Tillich et al.\ 2007).

The cross-correlation of this series of averaged spectra with a pure
hydrogen NLTE model spectrum at $T_{\rm eff} = 30\,000$\,K
and $\log{(g/\sf{cm\,s}^{-2})}=5.5$ as a template yields 
pulsational radial velocity measurements as shown in
Fig.~\ref{fig:schuh_fig1} for the different Balmer lines.
The maximum amplitude of a sinusoidal curve (fixed at the expected
period) that could be accommodated in comparison to the
weighted means of the Balmer lines reveals that any pulsational radial velocity amplitude
is smaller than the accuracy of our measurements and confirms the
upper limit of $16$\,{km\,s$^{-1}$} given by Kruspe et al.\ (2009).

The resolution of the model template matches the spectral resolution
of the (pulsation-averaged) Keck spectrum. A comparison of the
H$\alpha$ NLTE line core shape yields an even more stringent upper
limit for the combined broadening effect of pulsation and rotation of
at most $9$\,{km\,s$^{-1}$}.
This means that much better data in terms of spectral resolution and signal
to noise will be necessary to measure $v_{puls}$  and
$v_{rot}\sin{i}$. 

\newpage
 
\Acknowledgments{The authors thank A.~Reiners and G.~Basri for providing the Keck
  spectrum, H.~Edelmann for assistance in obtaining the HET
  spectra, and U.~Heber and T.~Rauch for providing
  grids of model spectra.
  This work has benefited from the help, advice and software by
  I.~Traulsen.
  We also thank the conference sponsors and in
  particular HELAS (European Helio- and Asteroseismology Network, an
  European initiative funded by the European Commission since April 1st,
  2006, as a ''Co-ordination Action'' under its Sixth Framework
  Programme FP6) and the Astronomische Gesellschaft for financially supporting
  the poster presentation at JENAM 2008 Minisymposium
  N$^{\circ}$~4 through travel grants to S.S.\ and R.L.
}

\References{
\rfr Kruspe, R., Schuh, S., Silvotti, R., \& Traulsen, I.\ 2008,
     CoAst, 157, 325
\rfr Lutz, R., Schuh, S., Silvotti, R., et al.\ 2008, ASP Conf.\ Ser., 392, 339
\rfr Lutz, R., Schuh, S., Silvotti, R., et al.\ 2009, A\&A, in press, arXiv:0901.4523
\rfr {\O}stensen, R., Solheim, J.-E., Heber, U., et al.\ 2001, A\&A, 368, 175
\rfr Silvotti, R., Janulis, R., Schuh, S., et al.\ 2002, A\&A, 389, 180
\rfr Silvotti, R., Schuh, S., Janulis, R., et al.\ 2007, Nature, 449, 189
\rfr Tillich, A., Heber, U., O'Toole, S.~J., et al.\ 2007, A\&A, 473, 219 
}
\end{document}